\documentclass[twocolumn, tighten]{aastex61}
\usepackage{amsmath}

\shorttitle{Searching X-ray spectra for narrowband communication}
\shortauthors{Hippke \& Forgan}

\begin{document}
\title{Interstellar communication. VI. Searching X-ray spectra for narrowband communication}

\author{Michael Hippke}
\affiliation{Sonneberg Observatory, Sternwartestr. 32, 96515 Sonneberg, Germany}
\email{hippke@ifda.eu}

\author{Duncan H. Forgan}
\affiliation{SUPA, School of Physics \& Astronomy, University of St Andrews, North Haugh, St Andrews, Scotland, KY16 9SS, UK}
\affiliation{St Andrews Centre for Exoplanet Science, University of St Andrews, UK}

\begin{abstract}
We have previously argued that targeted interstellar communication has a physical optimum at narrowband X-ray wavelengths $\lambda\approx1$\,nm, limited by the surface roughness of focusing devices at the atomic level \citep{2017arXiv171105761H}. We search $24{,}247$ archival X-ray spectra (of $6{,}454$ unique objects) for such features and present 19 sources with monochromatic signals. Close examination reveals that these are most likely of natural origin. The ratio of artificial to natural sources must be $\lesssim0.01$\%. This first limit can be improved in future X-ray surveys.
\end{abstract}

\section{Introduction}
Our previous work found the optimal frequency for data rate maximizing interstellar communication, given advanced technology, to be limited by the surface roughness of focusing devices at the atomic level. Depending on the material used for the reflective coating, the optimal wavelength is $\lambda\approx0.5-2$\,nm ($E\approx\,$keV) for distances out to kpc \citep{2017arXiv171105761H}. While this limit can be surpassed by beam-forming with electromagnetic fields (e.g. using a free electron laser), such methods are not energetically competitive. Current lasers are not yet cost efficient for nm wavelength, with a gap of two orders of magnitude, but future technological progress may converge on the physical optimum.

As detailed in \citet{2017arXiv171105761H}, the ideal spectrum for a maximum data rate connection will have a hard cut at $\lambda>0.5$\,nm due to mirror surface roughness. Bandwidth depends on the trade-off between the number of modes and the beam angle width. More wavelengths encode more bits per photon, however longer wavelengths have larger angular spread. The encoding efficiency follows a logarithmic relation with the number of modes \citep{2017arXiv171205682H}, so that the bandwidth will be small ($<100$\%) in realistic cases. As nanosecond time slots give $10^9$ modes per second, the monochromatic number of photons can be up to $10^8{\rm \,s}^{-1}$ before the mode penalty exceeds 1\% in bit rate. Consequently, a GB/s connection will be monochromatic if nanosecond technology is available. Such a connection over pc distances can be achieved with aperture sizes $D_{\rm t}=D_{\rm r}=10\,$m at modest (MW) power.

Optimal communication will be tightly focused with beam angles at the optimum wavelength $\lambda\approx$\,nm of $\theta=0.2\,{\rm mas}/D_{\rm t}$\,(m) due to diffraction. Randomly intercepting such beams between a large ($n=10,000$) club of communicating galactic civilizations is unlikely \citep{2014JBIS...67..232F}. Thus, we may only hope that such communication is directed at Earth.

Based on these assumptions, we may search the sky for narrowband or monochromatic sources between 0.5 and a few nm. A comparable search for optical laser signals in spectra was carried out by \citet{2017AJ....153..251T} with a null result for $5{,}600$ FGKM stars. There are $5.7\times10^7$ X-ray sources visible to XMM-Newton over the entire sky \citep{2007arXiv0704.2293C}, and spectra have been taken for $24{,}247$ ($4\times10^{-4}$ of the visible sources). Checking these spectra for artificial features can place a first upper bound to the number of optimally communicating (towards us) civilizations.

\begin{figure*}
\includegraphics[width=.5\linewidth]{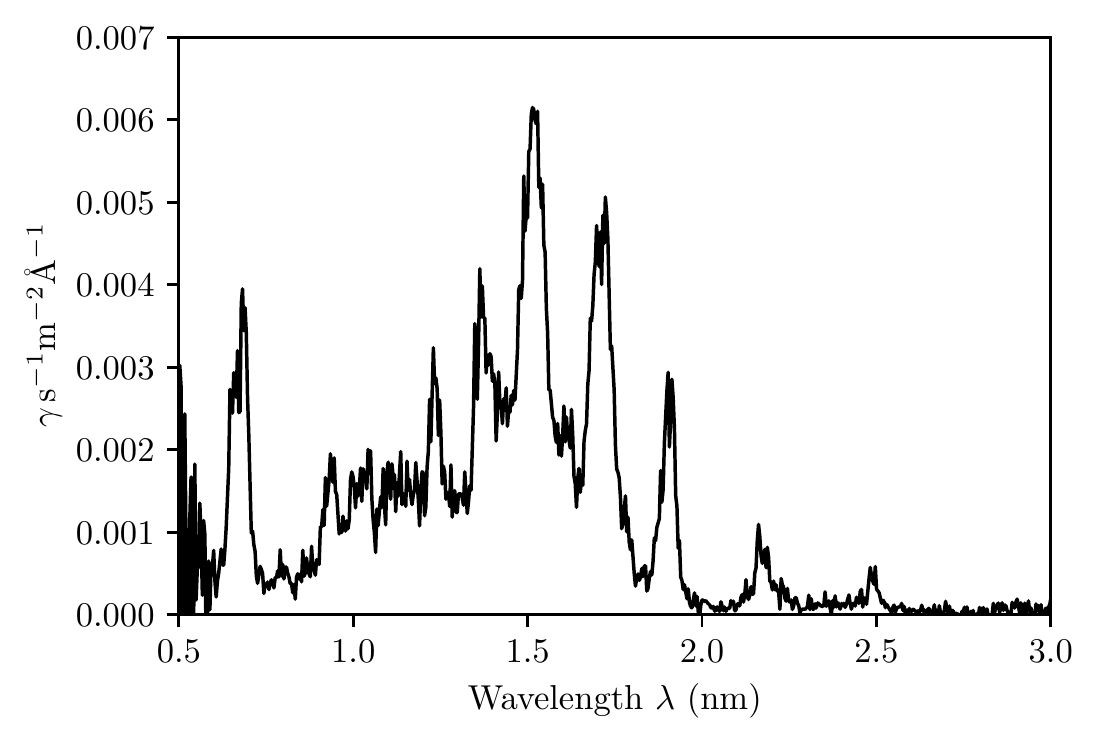}
\includegraphics[width=.5\linewidth]{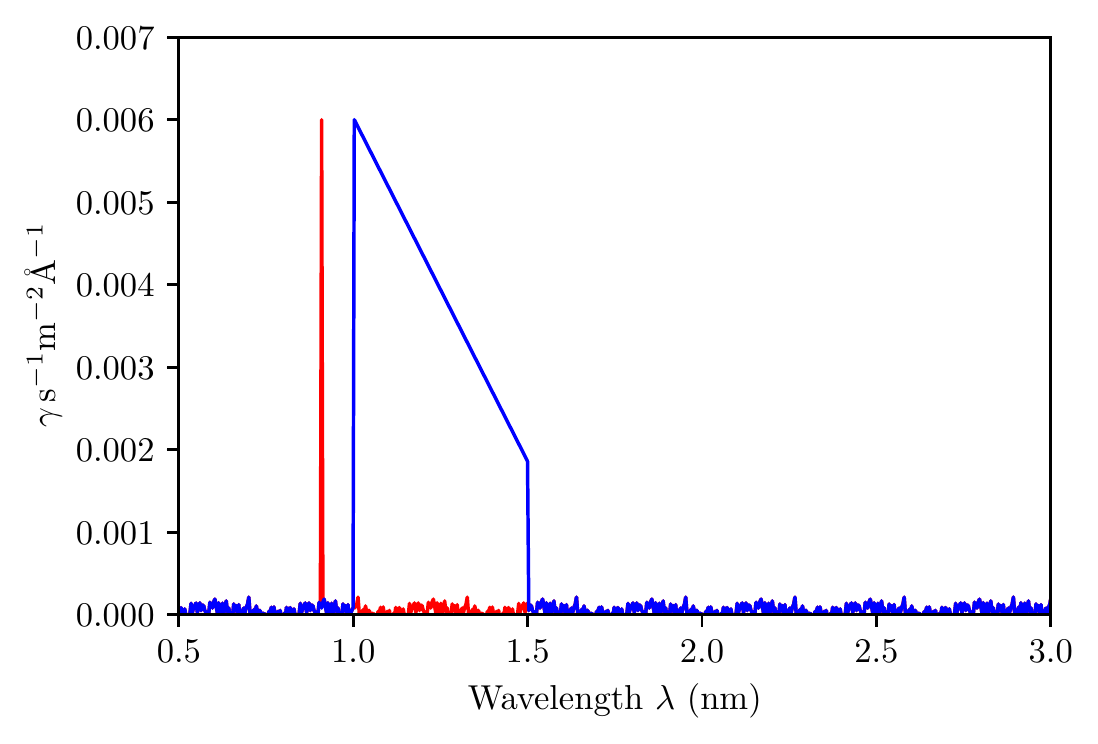}
\caption{\label{fig_xmm}Left: X-ray spectrum taken with XMM-Newton for Kepler's 1604 supernova remnant \citep{2004A&A...414..545C}. Right: Hypothetical spectra of monochromatic (red) and narrow (50\% bandwidth, blue) X-ray signals.}
\end{figure*}

\section{Method}
X-rays can only be observed from space due to atmospheric absorption. Several current X-ray satellites cover the wavelengths of interest:
Swift \citep[0.2--10\,keV,][]{2005SSRv..120..165B},
Chandra \citep[0.1--10\,keV,][]{2003SPIE.4851...28G} and
XMM-Newton \citep[0.1--12\,keV,][]{2001A&A...365L...1J}, while
INTEGRAL is only sensitive to higher energies (3\,keV--10\,MeV), and its spectrometer only covers the region 20 keV -- 8 MeV which is outside of our window of interest.
Because of their designs with grazing incidence mirrors, all telescopes have small collecting areas (0.01\,m$^2$, 0.04\,m$^2$ and 0.45\,m$^2$ for Swift, Chandra and XMM, respectively) \citep{2007A&A...463...79G}. With a spectral resolution of $R=800$ between 0.35--2.5\,keV (3.5\,eV at 1\,keV energy), XMM is well equipped for the proposed narrow-band X-ray signals \citep{2001A&A...365L...7D}.

As an example, we show the spectrum of Kepler's supernova remnant (SN1604) as observed with XMM-Newton by \citet{2004A&A...414..545C} in Figure~\ref{fig_xmm} (left panel). For comparison, synthetic monochromatic and narrow-band spectra are shown in the right panel.

\subsection{Data}
The latest generation of space-based X-ray telescopes offers high sensitivity, spatial resolution and energy range. Results demonstrate the potential of Galactic X-ray surveys to detect a wide variety of source types, such as coronally-active binaries \citep{2006ESASP.604...91H}, evolved protostars and T Tauri stars \citep{1999ARA&A..37..363F}, or isolated neutron stars \citep{2007A&A...476..317H}.

In this paper, we utilize observations drawn from the XMM-Newton public data archive. Specifically, we query the archive for all reduced, calibrated observations taken with the Reflection Grating Spectrometers which operate between $0.5-3.5$\,nm. This search yields a total of $24{,}247$ spectra of $6{,}454$ unique objects with exposure times between one second and 40 hours, and a mean (median) exposure time of 8.3 (5.8) hours. We download all spectra in FITS format using a custom-made script and create figures of each spectrum.

\subsection{Feature search}
The most prominent X-ray line emitting elements in the XMM-Newton passband are Iron, Oxygen, Magnesium, Sulfur, Silicon, Sodium, Calcium, Argon, Neon and Nickel \citep{2001A&A...365L...7D,2016A&A...594A..78G}. Most, but not all emission lines appear spectrally resolved \citep[e.g.,][]{2006A&A...449..475W,2016A&A...595A..85W}. Consequently, it is possible that natural emission resembles a single monochromatic signal, in case only one such narrow line emitting element would be detectable above the noise. We search for single peaks using a 1D wavelet \citep{doi:10.1093/bioinformatics/btl355} with accepted peak widths of 1--2 bins ($\approx3.5-7$\,eV at keV), as implemented by \texttt{SciPy}, and requiring SNR$>5$ for the highest peak, with no other peaks of SNR$>4$ being present in the same spectra. This criteria is necessary because many spectra show tens to hundreds of emission lines of natural origin, and it is impossible to distinguish these from the artificial case.

For the narrowband feature search, we required the presence of a peak with a minimum width of two bins ($\approx7$\,eV at keV) at SNR$>5$ and at least one sharp edge, i.e. a transition to SNR$<3$ on one side of the peak. No additional peaks with SNR$>4$ in the same spectrum were tolerated.

\begin{figure*}
\includegraphics[width=.5\linewidth]{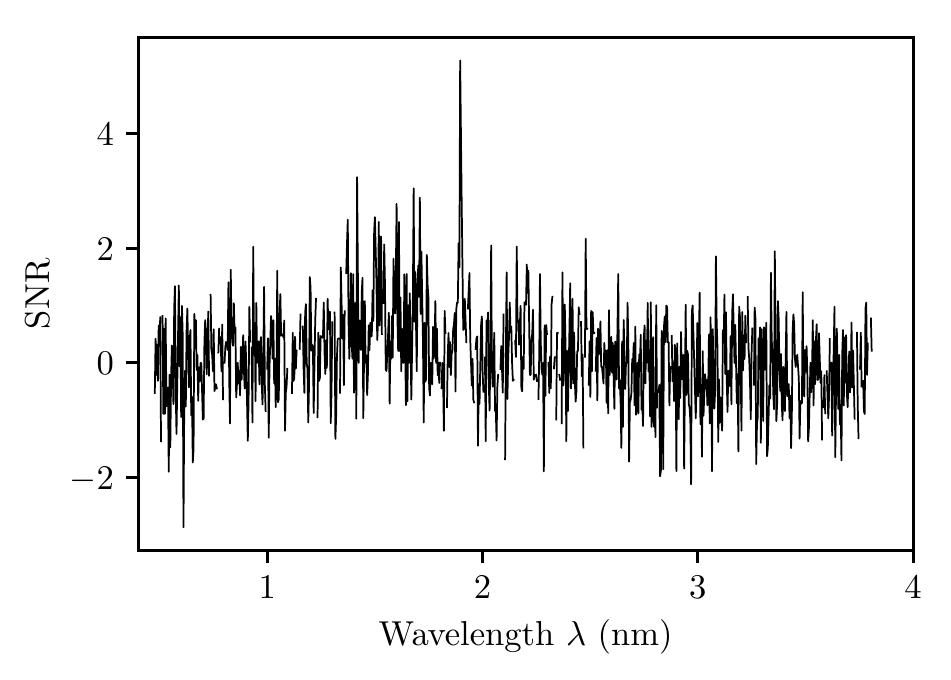}
\includegraphics[width=.5\linewidth]{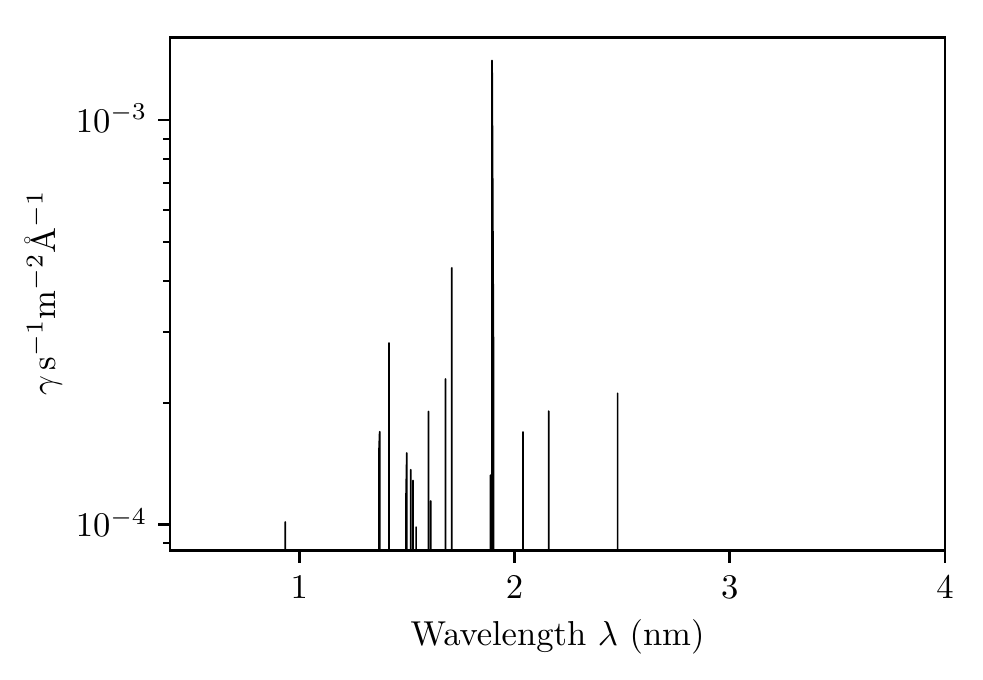}
\caption{\label{fig_result}Left: Spectrum for HD189733 in units of SNR used for visual examination (XMM-Newton ID\#0692290301). Only one peak with SNR$>5$ is found. Right: The same spectrum in physical flux units. For clarity, only channels with SNR$>2$ are shown.}
\end{figure*}

\section{Results}
After the automatic search and and visual examination, 19 spectra with a ``potentially monochromatic'' feature were identified (Table~\ref{table:mono}), and none with a narrowband feature. To verify the correctness of the algorithm, we relaxed the SNR requirements and visually examined the best candidates. No spectra of interest were found.

A literature search for the 19 spectra with a ``potentially monochromatic'' feature shows that all of them are identical with natural, astrophysical sources, such as galaxies and variable stars. Spatially resolved sources, such as the Cygnus Loop supernova remnant, exhibit different spectral features in different regions (pointings), resulting in single peak detection if SNR is low. The same mechanism appears to be present as well for all other sources. We found no indication for anything artificial.

As an example, rho Oph (the ``X-ray lighthouse'') shows strong and time-variable emissions between 0.5-3.5\,nm, including FeXXIV, NeX, NeIX, OVIII, and CVI \citep{2017AA...602A..92P}.

A spectrum for HD189733 is shown in Figure~\ref{fig_result}, with a prominent peak at 1.89\,nm, the oxygen emission line OVIII Lyman-$\alpha$ \citep{1974ApJ...192..169W,2011MNRAS.413.1251P}. This is the most commonly found emission line, detected in 7 of the 19 spectra. A detailed discussion of the natural emission properties of these lines is beyond the scope of this paper.

\section{Discussion}
We can calculate the expected SNR for a detector such as XMM-Newton for the case of optimal X-ray communication whose photon count scales as \citep{2017arXiv171105761H}

\begin{equation}
\gamma \approx d^{-2} D_{\rm t}^2 D_{\rm r}^2 P_{\rm t} {\rm (s^{-1})}
\end{equation}

where the distance $d$ is in pc, transmitter and receiver apertures $D_{\rm t}$ and $D_{\rm r}$ are in m and power $P_{\rm t}$ is in Watt. As an example, we use a $D_{\rm t}=1\,$m telescope at Proxima Cen ($d=1.3\,$pc) with $P=1\,$W and XMM-Newtons quantum efficiency of 50\% at keV energy \citep{2001A&A...365L..18S}. The monochromatic photon count for XMM is then 0.3\,\,s$^{-1}$, so that a high SNR detection with 100 photons in one spectral channel is achieved after 6\,min. 

If a survey spends 10\,min on each source, a clear detection in this framework is possible if the transmitter power $P_{\rm t} > 0.5 d^{2} D_{\rm t}^{-2}$. 

As an example, for $D_{\rm t}=1\,$m, $P_{\rm t}$ scales out to 50\,W (5\,kW, 500\,kW) at distances of 10\,pc (100\,pc, kpc). Equally, for kW power the aperture $D_{\rm t}$ would need to grow to 0.2 (2, 20\,kW) for distances of 10\,pc (100\,pc, kpc). 

These requirements are modest even for current technology, and it makes such communications technologically plausible. A survey with a duty cycle of 50\% using XMM-Newton could observe $5\times10^4$ sources per year. If the target list are the nearest stars of stellar types M, K, G, the survey could probe all of these stars out to 40 (100\,pc) within one (ten) years.
\\

\subsection{Upper limit on X-ray sources targeted at us}
The archival sample contains mostly established astrophysical sources, such as supernova remnants and radio-loud stars and galaxies. Targeting obviously natural (and sometimes hostile) sources makes the detection of artificial signals less likely, and upper limits too conservative. 

We detect no artificial signals in $6{,}454$ unique objects, and therefore the ratio of artificial to natural sources must be $\lesssim0.01$\%. This is not a strict limit, because the data does not come from a contiguous survey. An unbiased future spectral X-ray survey should target unlikely sources of powerful narrowband X-rays, such as nearby and field F, G, K, M stars.

\section{Conclusion}
Close examination of $24,247$ archival X-ray spectra revealed 19 candidates with single monochromatic emission lines. All of these are found to be likely of natural origin. This first limit can be improved in future X-ray surveys.

%\acknowledgments
%\texttt{Acknowledgments}
%The author is thankful to Ducan Forgan who suggested to perform this archival search.

\appendix
%pagebreak
\vspace{4cm}
\begin{rotatetable}
\begin{deluxetable*}{|c|c|c|c|c|}
\tablecaption{Objects with monochromatic X-ray signals \label{table:mono}}
\tablehead{\colhead{Object} & \colhead{Type} & \colhead{Peak $\lambda$ (nm)} & \colhead{ID\#} & \colhead{Comment}}
\startdata
NGC 720             & Galaxy                   & 0.521 & 0112300101 & Other peaks exist \citep{2003ApJ...585..756J} \\
4C 29.30            & Seyfert 2 galaxy         & 3.851 & 0504120101 & Strong GHz source \citep{2007MNRAS.378..581J}\\
Cygnus Loop         & Supernova remnant        & 3.426 & 0018140101 & Extended source \citep{1997ApJ...484..304L}\\
\tableline
zet CMa             & Cepheid variable         & 1.899 & 0600530101 & Pulsating magnetosphere \citep{2017MNRAS.471.2286S}\\
HD 189733           & BY Dra type variable     & 1.896 & 0744981701 & Hot Jupiter exoplanet \citep{2005AA...444L..15B}\\
eps Per             & Beta Cep type variable   & 1.896 & 0761090801 & Binary \citep{2006AA...446..583L}, pulsations \citep{2000ApJ...543..359S} \\
Kappa Sco           & Beta Cep type variable   & 2.162 & 0503500201 & Binary, pulsations \citep{2004AA...422.1013H,2005AA...432..955U}\\
56 Tau              & Alpha2 CVn type variable & 1.896 & 0201360201 & Period 1.57\,d \citep{1995AAS..111...41N}\\
HD 283572           & T Tau-type variable      & 0.524 & 0101441001 & SED peak at keV \citep{1998AA...337..413F}\\
CE 315 (V396 Hya)   & Nova, AM CVn binary      & 0.557 & 0302160201 & Cataclysmic variable \citep{2006AA...457..623R}\\
RBS 1955 (V405 Peg) & Dwarf nova               & 0.589 & 0604060101 & Cataclysmic binary \citep{2009PASP..121..465T}\\
61 Cyg A+B          & Double star              & 1.899 & 0041740301 & Nearby K5V+K7V \citep{2012AA...543A..84R,2016AA...594A..29B}\\
rho Oph             & Double star              & 0.553 & 0305540601 & ``X-ray lighthouse'' period 1.4\,d \citep{2017AA...602A..92P}\\
HD 54662             & Double star             & 0.575 & 0780150101 & Spectroscopic binary \citep{2007ApJ...664.1121B}\\
IGR J17091-3624	    & Low mass X-ray binary    & 0.560 & 0406140401 & ``Exotic variability'' \citep{2017MNRAS.468.4748C}\\
HD 79555            & High proper-motion star  & 1.896 & 0602830401 & Young \citep[35\,Myr,][]{2010AA...521A..12M} binary \citep{2001AJ....122.3466M}\\
CN Leo (Wolf 359)   & Flare star               & 1.899 & 0200530701 & 3rd closest star (2.4\,pc), dM6e \citep{1969PASP...81..527W}\\
UV Ceti	(GJ 65 B)   & Flare star               & 0.474 & 0111430201 & Nearby binary M5.5Ve+M6Ve \citep{2016AA...593A.127K}\\
eta Car             & Emission-line star       & 2.483 & 0560580201 & Colliding-wind binary \citep{2017arXiv170801033L}\\
\enddata
\end{deluxetable*}
\end{rotatetable}

\bibliographystyle{yahapj}

\end{document}